\documentstyle[12pt,epsf]{article}

\newcommand{\gapprox}{\raisebox{-.2ex}{$\stackrel{\textstyle>}
{\raisebox{-.6ex}[0ex][0ex]{$\sim$}}$}}

\begin{document}

\begin{flushright}
CEBAF-TH-95-17 \\
November 1995\\
hep-ph/9511270
\end{flushright}
\vspace{2cm}
\begin{center}
{\Large\bf QCD Sum Rule Calculation \\ of
$\gamma\gamma^*\to\pi^0$ Transition Form Factor
}
\end{center}
\begin{center}
{A.V.RADYUSHKIN\footnotemark
 \\
{\em Physics Department, Old Dominion University, Norfolk, VA 23529,USA} \\
{\em and}\\ {\em Continuous Electron Beam Accelerator Facility,} \\
 {\em Newport News, VA 23606, USA}\\ [2.0mm]
R.RUSKOV\footnotemark \\
{\em Laboratory of Theoretical Physics, JINR, Dubna, Russian Federation}\\ }
\end{center}
\footnotetext{Also at the Laboratory of Theoretical Physics,
JINR, Dubna, Russian Federation}
\footnotetext{On leave of absence from Institute of Nuclear
Research and Nuclear Energy,
Bulgarian Academy of Sciences, 1784, Sofia, Bulgaria}

\begin{abstract}

We develop  a QCD sum rule analysis of the form factor
\mbox{$F_{\gamma^*\gamma^*\pi^\circ}$}$(q^2,Q^2)$
in the region where  virtuality of one of the spacelike photons
is small
$q^2 \ll  1{\mbox{ GeV}}^2$ while  another is large:
 $Q^2 \gapprox \, 1{\mbox{ GeV}}^2$.
We construct  the operator product expansion
suitable for this   kinematic situation and
obtain  a   QCD sum rule for \mbox{$F_{\gamma^*\gamma^*\pi^\circ}$}$(0, Q^2)$.
Our results confirm expectation
that the momentum transfer dependence of
\mbox{$F_{\gamma^*\gamma^*\pi^\circ}$}$(0, Q^2)$
is close to interpolation
between its  $Q^2=0$ value fixed by the axial anomaly
and $Q^{-2}$ pQCD behaviour  for large $Q^2$.
 Our approach, in contrast to pQCD,
does not require  additional assumptions
about the shape of the pion distribution amplitude $\varphi_{\pi}(x)$.
 The  absolute value  of the
$1/Q^2$ term obtained in this paper
favours  $\varphi_{\pi}(x)$ close to
the asymptotic form  $\varphi_{\pi}^{as} (x)=6 f_{\pi} x (1-x)$.

\end{abstract}

\newpage

\section{Introduction.}

 The transition $\gamma^* \gamma^* \to \pi^0$
of two virtual photons $\gamma^*$ into a neutral pion
provides an exceptional opportunity to test
QCD predictions for exclusive processes.
In the lowest order of
perturbative QCD, its asymptotic behaviour
is due to the subprocess
$\gamma^*(q_1) + \gamma^*(q_2) \to \bar q(\bar xp) + q (xp) $
with $x$ ($\bar x$) being  the fraction of the pion momentum $p$ carried
by the quark produced at the $q_1$ ($q_2)$ photon vertex (see Fig.1$a$).
The relevant diagram is similar to the handbag
diagram for deep inelastic scattering,
with the main difference that one should use
 the pion distribution amplitude $\varphi_{\pi}(x)$
instead of parton densities.
For large $Q^2$, the perturbative
QCD prediction is given by \cite{bl80}:
\begin{equation}
F_{\gamma^* \gamma^*  \pi^0 }^{as}(q^2, Q^2) = \frac{4\pi}{3}
\int_0^1 {{\varphi_{\pi}(x)}\over{xQ^2+\bar x q^2}} \, dx
\stackrel{q^2=0}{\longrightarrow}
\frac{4\pi}{3}
\int_0^1 {{\varphi_{\pi}(x)}\over{xQ^2}} \, dx
\equiv \frac{4\pi}{3Q^2} I
\label{eq:gg*pipqcd}
\end{equation}
($Q^2   \equiv - q_2^2$, $q^2 \equiv -q_1^2$ and our convention
is $q^2 \leq Q^2$).
Experimentally,
the most important situation is when one of the photons
is almost real $q^2 \approx 0$. In this  case,  necessary
nonperturbative information
is accumulated   in
the same integral $I$  (see eq.(\ref{eq:gg*pipqcd}))
that appears  in the one-gluon-exchange
diagram  for the  pion electromagnetic  form factor
\cite{pl80,blpi79,cz84}.
The  value of $I$ depends on the shape of the
pion distribution amplitude $\varphi_{\pi}(x)$.
In particular,  using  the
asymptotic form
$
\varphi_{\pi}^{as}(x) = 6 f_{\pi} x \bar x
$
\cite{pl80,blpi79} gives $F_{\gamma \gamma^*  \pi^0 }^{as}(Q^2) =
4 \pi f_{\pi}/Q^2 $ for  the asymptotic
behaviour \cite{bl80}. If one takes the
Chernyak-Zhitnitsky form
$\varphi_{\pi}^{CZ}(x) = 30 f_{\pi} x(1-x)(1-2x)^2$ \cite{cz82},
the integral $I$ increases by a sizable factor of 5/3,
and this difference can be used for experimental
discrimination between the two forms.

An important point is that,
unlike the case of the pion EM form factor,
the pQCD  hard scattering term for $\gamma \gamma^* \to \pi^0$
($\gamma$ denoting a real photon) has  zeroth
order in the QCD coupling constant $\alpha_s$,
$i.e.,$  the
asymptotically leading term has no suppression.
The  situation is  similar   to that in deep inelastic
scattering. Hence, we have good reasons to
expect that pQCD   for $F_{\gamma \gamma^*  \pi^0 }(Q^2)$
 may work at accessible $Q^2$.
Of course, the asymptotic $1/Q^2$-behaviour
cannot be true in the  low-$Q^2$
region,  since  the $Q^2=0$ limit of
$F_{\gamma \gamma^*  \pi^0 }(Q^2)$
is known to be finite and
 normalized by the $\pi^0 \to \gamma \gamma$ decay rate.
Using  PCAC and ABJ anomaly \cite{ABJ},
one can calculate $F_{\gamma \gamma^*  \pi^0 }(0)$
theoretically:
$
F_{\gamma \gamma^*  \pi^0 }(0) =1/ \pi f_{\pi} .
$
It is natural to expect that a complete QCD result
does not  strongly deviate from
a simple interpolation
$
\pi f_{\pi} F_{\gamma \gamma^*  \pi^0 }(Q^2) =
1/(1+ Q^2/4 \pi^2 f_{\pi}^2)
$  \cite{blin}
between
the $Q^2=0$ value and the large-$Q^2$
asymptotics\footnote{In particular, such an
interpolation agrees with the results of  a
constituent quark model calculation \cite{hiroshi}}.
This interpolation implies the asymptotic form
of the distribution amplitude for the large-$Q^2$ limit and
agrees  with CELLO
experimental data \cite{CELLO}.
It was also claimed \cite{CLEO} that the new CLEO data
available  up to $8 \, GeV^2$  also
agree with the interpolation  formula.
This provides a strong evidence that the pion
distribution amplitude is  rather close to
 its asymptotic form.
Because of the far-reaching consequences   of this conclusion,
it is desirable  to have  a direct  QCD  calculation
of  the  $\gamma\gamma^* \to \pi^0$
form factor in the intermediate region of  moderately large
momentum transfers $Q^2 \gapprox 1 \, GeV^2$.
Such an approach is provided by
QCD sum rules. As we will see below,
 the QCD sum  rules  also allow one  to
calculate $F_{\gamma \gamma^*  \pi^0 }(Q^2)$  for large $Q^2$
without any  assumptions
about the shape of the pion distribution amplitude.
In fact, the QCD sum rule for $F_{\gamma \gamma^*  \pi^0 }(Q^2)$
can be used to get  information about  $\varphi_{\pi}^{as}(x)$.

\section{ Definitions.}

The $\gamma^*\gamma^* \to \pi^0$  transition
form factor $F_{\gamma^*\gamma^* \pi^0}(q^2,Q^2)$
  can be defined through  the matrix element
\begin{eqnarray}
\int
\langle {\pi},\stackrel{\rightarrow}{p}
|T\left\{J_{\mu }(X)\,J_{\nu}(0)\right\}| 0 \rangle e^{-iq_1 X } d^4 X
 = \alpha \sqrt{2} \epsilon_{\mu  \nu \alpha  \beta}
q_1^{\alpha} q_2^{\beta} \,\mbox{$F_{\gamma^*\gamma^*\pi^\circ}$}\left(q^2,Q^2
\right ) ,
\label{eq:form}
\end{eqnarray}
where
$\alpha = e^2/4 \pi$ is the fine structure  constant,
$
J_{\mu }={\it{e}}\left({2\over{3}}\bar{u}\gamma_{\mu }
u\,-\,{1\over{3}}\bar{d}\gamma_{\mu }d\right)\,
$
is the electromagnetic current of the light quarks
and  $|  {\pi} , \stackrel{\rightarrow}{p} \rangle$  is a $\pi^0$ state
with  the 4-momentum $p$.
To incorporate QCD sum rules \cite{svz}, we consider a
  three-point correlation function
\begin{equation}
{\cal{F}}_{\alpha\mu \nu}(q_1,q_2)= \frac{i} {\alpha \sqrt{2}}
\int
\langle 0 |T\left\{j_{\alpha}^5(Y) J_{\mu }(X)\,J_{\nu}(0)\right\}| 0 \rangle
e^{-iq_{1}X}\,e^{ipY}  d^4X\,d^4Y \,  ,
\label{eq:corr}
\end{equation}
 (cf. \cite{NeRa83})
containing  the axial current
$
j_{\alpha}^5 = {1\over{\sqrt{2}}}{\left(\bar{u}\gamma_{5}
\gamma_{\alpha}u\,-\,\bar{d}\gamma_{5}\gamma_{\alpha}d\right)}
$
serving as a field with a non-zero
projection onto the neutral pion state:
$
\langle 0 |j_{\alpha}^5(0)| {\pi^0},\stackrel{\rightarrow}{p}\rangle =
 -i\,f_{\pi} p_{\alpha}.
$
The three-point amplitude ${\cal{F}}_{\alpha\mu \nu}(q_1,q_2)$
has a  pole for $p^2=m_{\pi}^2$:
\begin{equation}
{\cal{F}}_{\alpha\mu \nu}(q_1,q_2) =
\frac{f_{\pi}}{p^2-m_{\pi}^2}
p_{\alpha}  \epsilon_{\mu  \nu \alpha  \beta}
q_1^{\alpha} q_2^{\beta}
\mbox{$F_{\gamma^*\gamma^*\pi^\circ}$} (q^2,Q^2)+ \ldots \  ,
\label{eq:pole}
\end{equation}
$i.e.,$ the Lorentz
structure of  the pion contribution
is  $p_{\alpha}  \epsilon_{\mu  \nu \alpha  \beta}
q_1^{\alpha} q_2^{\beta} $,
and  the spectral density of the  dispersion relation
 \begin{equation}
{\cal{F}}\left(p^2,q^2,Q^2\right)={1\over{\pi}}\int_0^{\infty}
\frac{{\rho}\left(s,q^2,Q^2\right)}{s-p^2}\,ds
+ ``subtractions" .
\label{eq:disp1}
\end{equation}
for the relevant invariant amplitude   can be written as
\begin{equation}
\rho \left(s,q^2,Q^2\right)=
\pi f_{\pi}\delta(s-m_\pi^2)
\mbox{$F_{\gamma^*\gamma^*\pi^\circ}$}\left(q^2,Q^2\right)
+  ``higher \ states".
\label{eq:rho}
\end{equation}
The higher states  include $A_1$ and higher  broad
pseudovector resonances. Due to asymptotic freedom,
their sum, for large $s$, rapidly approaches the
pQCD  spectral density ${\rho}^{PT}(s,q^2,Q^2)$.
The simplest model is to approximate all the higher states,
including the $A_1$, by the
perturbative contribution:
\begin{equation}
\rho^{mod}\left(s,q^2,Q^2\right) =
\pi f_{\pi}\delta(s)\mbox{$F_{\gamma^*\gamma^*\pi^\circ}$}\left(q^2,Q^2\right)
+\theta(s-s_o){\rho}^{PT}(s,q^2,Q^2)
\label{eq:rhoph}
\end{equation}
where the parameter $s_o$ is the effective threshold for   higher states.
To suppress the higher states by an
exponential weight $\exp[-s/M^2]$,
 we  apply  the SVZ-Borel transformation \cite{svz}:
\begin{equation}
\hat B(-p^2\rightarrow M^2){\cal F}(p^2, q^2,Q^2)
\equiv \Phi (M^2,q^2,Q^2)
=\frac{1}{\pi M^2} \int_0^\infty  e^{-s/M^2} {\rho}(s,q^2,Q^2) \, ds,
\label{eq:fph}
\end{equation}
which, moreover,   produces
a  factorially improved OPE power series for large $M^2$:
$1/(-p^2)^N \to (1/M^2)^N / (N-1)!$.

\begin{figure}[t]
\mbox{
   \epsfxsize=12cm
 \epsfysize=4cm
  \epsffile{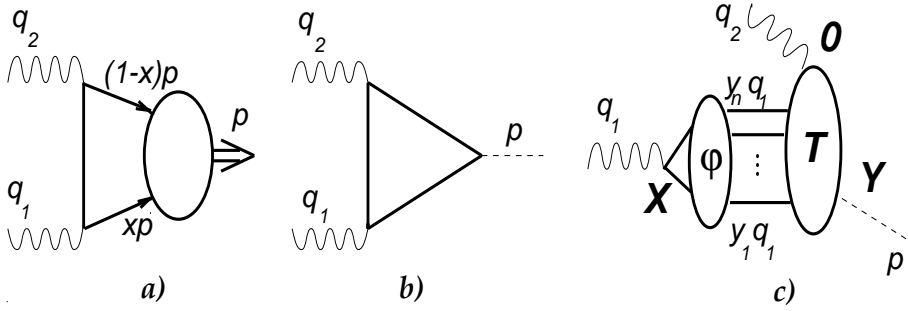}  }
  \vspace{1cm}
{\caption{\label{fig:1}
{\it a)} Leading-order pQCD contribution.
{\it b)} Triangle diagram. {\it c)} General structure of bilocal
contributions.
   }}
\end{figure}

To construct   a QCD sum rule, one  should  calculate
the three-point function ${\cal F}(p^2,q^2,Q^2)$
and then its SVZ-transform   $\Phi (M^2,q^2,Q^2)$
as a  power  expansion in $1/M^2$  for large $M^2$.
However, a particular   form of the  original
$(1/p^2)^N$-expansion depends
on the   values of the photon virtualities
$q^2 $ and  $Q^2$.

\section { QCD sum rules for  large $q^2$.}

The simplest    case  is when both
virtualities are  large:
$Q^2 \sim q^2 \sim - p^2 > \mu ^2$ where $\mu $ is some
scale  $\mu ^2 \sim 1 \, GeV^2$
above which one can rely on  asymptotic freedom
of QCD.
 Then all
  contributions which have  a power behaviour $(1/M^2)^N$
correspond to a  situation  with all three currents
close to each other:    all the
  intervals $X^2, Y^2,(X-Y)^2$\, are small.

The starting point   of the OPE  is  the
perturbative triangle graph (Fig.\ref{fig:1}$b$).
Using  Feynman
parameterization and performing simple integrations  we get:
\begin{equation}
\rho^{PT}(s,q^2,Q^2)=2\int_0^1 \frac{x\bar{x}(xQ^2+ \bar x q^2)^2}
{[s{x}\bar{x}+xQ^2+ \bar x q^2]^3} \,dx  \, .
\label{eq:rhopt}
\end{equation}
The variable $x$
is the light-cone fraction of the total pion
momentum $p$ carried by one of the quarks.
Adding the condensate corrections,
we obtain the  following QCD sum rule:
\begin{eqnarray}
\pi f_{\pi} \mbox{$F_{\gamma^*\gamma^*\pi^\circ}$}(q^2,Q^2)=
 2\int_0^{s_o} ds \, e^{-s/{M^2}}
\int_0^1 \frac{x\bar{x}(xQ^2+ \bar x q^2)^2}
{[s{x}\bar{x}+xQ^2+ \bar x q^2]^3} \,dx  \,
\nonumber \\
+\frac{\pi^2}{9}
{\langle \frac{\alpha_s}{\pi}GG \rangle}
\left(\frac{1}{2M^2 Q^2} + \frac{1}{2M^2 q^2}
 - \frac1{Q^2 q^2}\right)
 \nonumber\\
+ \frac{64}{243}\pi^3\alpha_s{\langle \bar{q}q\rangle}^2
\left( \frac1{M^4}
\left [ \frac{Q^2}{q^4}+ \frac9{2q^2}+\frac9{2Q^2}+\frac{q^2}{Q^4} \right ] +
\frac9{Q^2 q^4} +\frac9{Q^4 q^2} \right )  .
\label{eq:SR1}
\end{eqnarray}
It is valid when  both
virtualities of the photons are large.
In this  region,  the  perturbative QCD approach
is also expected to work.
This expectation is completely
supported by our sum rule.
Indeed,  neglecting  the $s{x}\bar{x}$-term
compared to $xQ^2+ \bar x q^2$
and keeping only the leading  $O(1/Q^2)$
and $O(1/q^2)$ terms in the condensates,
 we can write eq.(\ref{eq:SR1}) as
\begin{eqnarray}
\mbox{$F_{\gamma^*\gamma^*\pi^\circ}$}(q^2,Q^2)
= \frac{4\pi}{3f_{\pi}}
   \int_0^1 \frac{dx}{ ( xQ^2 + \bar x q^2)} \,
\left \{ \frac{3M^2}{2\pi^2}(1-e^{-s_0/M^2}) x\bar{x}
\right. \nonumber \\ \left.
+ \frac{1}{24M^2}
\langle \frac{\alpha_s}{\pi}GG\rangle [\delta(x) + \delta (\bar{x})]
\right. \nonumber \\
+ \left. \frac{8}{81M^4}\pi\alpha_s{\langle \bar{q}q\rangle}^2
 \biggl ( 11[\delta(x) + \delta (\bar{x})] +
2[\delta^{\prime}(x) + \delta ^{\prime}(\bar{x})]
\biggr ) \right \}
\label{eq:SRlargeQ2wf}.
\end{eqnarray}
The expression in  curly brackets
coincides with the QCD sum rule for
the pion distribution amplitude
$f_{\pi} \varphi_{\pi}(x)$ (see, $e.g.,$ \cite{mr}).
Hence, when  $q^2$, the smaller of two  photon virtualities
is large,
the QCD sum rule
(\ref{eq:SR1})
exactly reproduces  the pQCD result (\ref{eq:gg*pipqcd}).

\section { Operator product expansion for small $q^2$.}

One may be tempted to  get
a QCD sum rule for the integral $I$ by taking  $q^2=0$
in eq.(\ref{eq:SR1}).
Such an attempt, however, fails immediately
because of the power singularities
$1/q^2$, $1/q^4$, $etc.$
in the condensate terms.
It is easy to see that these singularities
are produced by  the
$\delta(x)$ and $\delta'(x)$ terms in eq.(\ref{eq:SRlargeQ2wf}).
In fact, it is precisely these terms that  generate the two-hump form
for $\varphi_{\pi}(x)$ in the CZ-approach \cite{cz82}.
Higher condensates would produce even more singular
$\delta^{(n)}(x)$ terms.  As shown in ref.\cite{mr},
the  $\delta^{(n)}(x)$ terms result from the Taylor expansion of
nonlocal condensates like $\langle \bar q(0) q(Z) \rangle$.
Modelling  nonlocal condensates by functions decreasing
at large $(-Z^2)$, $i.e.,$ assuming a finite correlation
length $\sim 1/\mu$ for vacuum fluctuations,
one obtains smooth curves instead of  the singular
$\delta(x)$ and $\delta'(x)$ contributions, and the  result for
$\varphi_{\pi}(x)$ is close to the asymptotic form \cite{mr}.
Effectively, the correlation length provides
an IR cut-off in the end-point regions $x\sim 0, x \sim 1$.
Similarly,  we expect that, when
$q^2$ is too small to provide  an appropriate IR cut-off
in the sum rule (\ref{eq:SR1}),
such a cut-off is again generated  by  nonperturbative effects,
$i.e.,$ that
 eventually  $1/q^2$
is  substituted  for small $q^2$ by something
like $1/m_{\rho}^2$.  Below, we show
that this is exactly what happens in the QCD sum rule framework.

To illustrate the nature of the modifications
required in the small-$q^2$ region,
 it is instructive to analyze first
the perturbative term. The  latter,
though finite for    $q^2 = 0$,
contains contributions which
are  non-analytic at this point:
\begin{eqnarray}
\Phi^{PT}(q^2,Q^2,M^2)=
\frac1{\pi M^2} \int_0^{\infty} e^{-s/M^2}
\left\{ \ 1+
 \left[2\frac{q^2y}{M^2}  \hspace{2cm}
\right. \right.
\nonumber \\
\left. \left.
+
\frac{q^4y^2}{M^4}\right] e^{{q^2y}/{M^2}} \ln{\left
( \frac{q^2y}{M^2}\right )}   + \ldots \right \}
\frac{Q^2  ds}{(s+Q^2)^2},
\label{eq:phipt}
\end{eqnarray}
where $y=s/(s+Q^2)$ and dots
stand for  terms analytic and  vanishing for $q^2=0$.
The logarithms here  are a typical example
of mass singularities (see, $e.g.,$ \cite{georgietal,sterman}
and, for QCD sum rule applications, refs.  \cite{tkachev,nr84,BelKog}):
the singularities  are   due to the possibility of the
long-distance propagation in the $q$-channel.
In  other words, when $q^2$ is small,
 there appears an additional
possibility to get a power-behaved contribution
from a configuration  in which
the large momentum   flows
from  the $Q$-vertex into the  $p$-vertex
(or $vice$ $versa$)
without entering  the $q$-vertex,
and small momenta flowing through other parts
of the diagrams induce singular contributions.
In the coordinate representation, such a configuration
can be realized by keeping
the electromagnetic current $J_{\mu }(X)$ of
 the low-virtuality photon
far away from  two other currents,
which are still close to each other.
The contribution
generated in  this regime can be extracted
 through  an  operator product expansion
for the short-distance-separated currents:
 $T\{ J(0) j^5(Y) \}  \sim  \sum C_i(Y)
{\cal O}_i. $
Diagrammatically, the situation is similar to the pQCD limit
$q^2,Q^2 \gg -p^2$ discussed above.
The only difference is that we should consider now the limit
$-p^2,Q^2 \gg q^2$.
The result again can  be written
in a ``parton'' form (see Fig.1$c$):
\begin{equation}
{\cal F}^{bilocal}(q_1,q_2,p) =
\int_0^1 \phi_{\gamma}^{(i)} (\{y\},q^2) T^{(i)}
(\{yq_1\};q_2,p) [dy],
\label{eq:Fbilocal}
\end{equation}
where $\phi_{\gamma}^{(i)}  (\{y\},q^2)$ can be treated as  distribution
amplitudes of the $q_1$-photon,
with $y$'s  being the light-cone fractions
of the  momentum  $q_1$ carried by  the relevant partons
($i.e.,$ quark and gluonic $G$-fields present
in ${\cal O}$).
The functions $\phi_{\gamma}^{(i)}  (\{y\},q^2)$
 are related to  the
correlators (``bilocals'', cf. \cite{balyung})
\begin{equation}
\Pi^{(i)}(q_1) \sim \int e^{iq_1 X} \langle 0| T \{ J_{\mu }(X)
{\cal O}^{(i)}  (0) \}| 0 \rangle d^4X
\label{eq:bilocal}
\end{equation}
of the $J_{\mu }(X)$-current with  composite operators
$ {\cal O}^{(i)}  (0)$.
Performing  such a factorization
for each diagram,   we represent the
amplitude ${\cal F}$   as a  sum of its
purely  short-distance ($SD$) and bilocal ($B$) parts.
The $SD$-part (which is  defined
as the difference between the original unfactorized expression
and the perturbative version  of
its $B$-part)
is regular in the $q^2 \to 0$ limit and can be  treated perturbatively.
On the other hand, the low-$q^2$ behaviour of the
$B$-correlators $\Pi(q_1)$ cannot be directly
calculated  in  perturbation theory.

\begin{figure}[t]
\mbox{
   \epsfxsize=12cm
 \epsfysize=4cm
  \epsffile{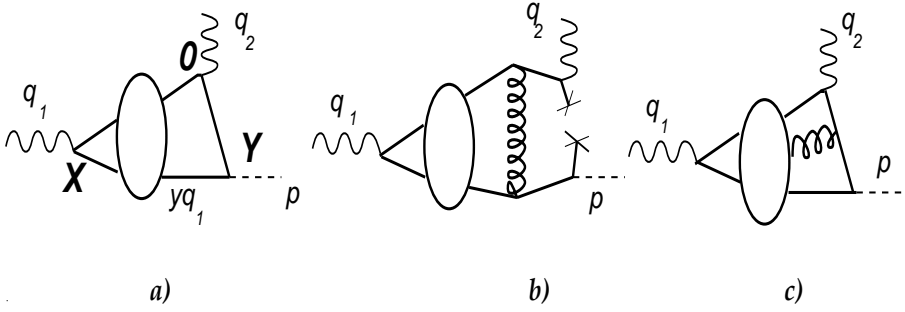}  }
  \vspace{1cm}
{\caption{\label{fig:2}
Bilocal contributions with coefficient functions
given by
{\it a)} single propagator;
{\it b)} product of 3 propagators; {\it c)} product of 2 propagators.
   }}
\end{figure}

\section{ Structure of bilocal contributions.}

In the simplest case, the amplitude
$ T^{(i)} (\{yq_1\};q_2,p)$  in the
bilocal term is given by a single quark propagator (Fig.2$a$):
\begin{equation}
T(yq_1, \bar y q_1;q_2,p)  \sim \frac1{(p-yq_1)^2}=
\frac1{\bar y p^2 + y \bar y q^2 -y Q^2},
\label{eq:sinprop}
\end{equation}
accompanied by  two-body distribution
amplitudes $\phi_{\gamma}^{(i)}  (y,q^2)$,
with $yq_1$ and $\bar y q_1 $ being the
momenta carried by the quarks ($\bar y\equiv 1-y$).
The $y^n$-moments of $\phi_{\gamma}^{(i)}  (y,q^2)$
are given by bilocals $\Pi_n^{(i)}(q^2)$ involving
composite operators ${\cal O}_n^{(i)}$ with
$n$ covariant derivatives.
Note, that it is legitimate to keep the
$q^2$-term in eq.(\ref{eq:sinprop})
when substituting it into eq.(\ref{eq:Fbilocal}):
all the $(q^2/Q^2)^N$ and $(q^2/p^2)^N$
power corrections  are exactly reproduced  there
due to a phenomenon
analogous to  the $\xi$-scaling
\cite{xiscaling} in deep inelastic scattering.
Applying the SVZ-Borel transformation, we obtain the result
\begin{equation}
\Phi^{B}_1(q^2,Q^2,M^2) \sim \frac1{M^2}
\int_0^1 \phi_{\gamma} (y,q^2)
e^{y q^2 / M^2}
e^{- y Q^2 / \bar y  M^2} dy ,
\label{eq:phising}
\end{equation}
which has the structure of eq.(\ref{eq:phipt}): one should just
take  $y Q^2 /\bar y =s$.

In perturbation theory, the amplitudes $\phi_{\gamma}^{(i)} (y,q^2)$
have logarithmic non-analytic behaviour for
$q^2=0$.
Eq.(\ref{eq:phipt}) indicates  that,
for the triangle graph, there are only
two independent sources of
logarithmic singularities.
They correspond to two-body operators of  leading
and next-to-leading twist in the OPE for $T\{J(0)j_5(Y)\}$.
The $\log q^2 $-terms
 reflect the fact that
 the lowest singularity  in
$\Phi^{PT}(q^2,Q^2,M^2)$
for the $q$-channel corresponds to  threshold
of the $q\bar{q}$-pair production which,
for massless quarks, starts  at zero.
However,    this is true  only in
perturbation theory.
For hadrons, the first singularity
in the $q$-channel is located at the  $\pi\pi$
threshold, with the $\rho$-resonance
being the most prominent feature of the physical
spectral density for the correlators $\Pi_n^{(i)}(q^2)$ .
In other words, the  $1/q^2$ and $1/q^4$ terms in the condensates
and logarithmic  terms in the
perturbative contribution
correspond to the  operator product
 expansion for the correlators $\Pi_n^{(i)}(q^2)$,
which is  only valid
in the large-$q^2$ region.
To get  $\Pi_n^{(i)}(q^2)$  for small $q^2$,
one can  use this large-$q^2$ information to
construct a model for the physical spectral density
$\sigma_n^{(i)}(t)$  and then calculate $\Pi_n^{(i)}(q^2)$
from the dispersion relation
\begin{equation}
\Pi_n^{(i)}(q^2) = \frac1{\pi} \int_0^{\infty}
\frac{\sigma_n^{(i)}(t) dt}{t+q^2}.
\label{eq:pidr}
\end{equation}
To this end, we use the usual ``first resonance plus continuum''
model
\begin{equation}
\sigma_n^{(i)}(t) = g_{n}^{(i)} \delta(t-m_{\rho}^2) +
\theta(t> s_{\rho})\sigma_n^{(i)PT}(t) .
\label{eq:sigma}
\end{equation}
Practically, this means that we
modify the original spectral density in the
region $t< s_{\rho}$ by
subtracting  all the terms of the OPE
for $\sigma_n(t)$ in this region
and replace them  by the $\rho$-meson
contribution  $g_{n} ^{(i)}\delta(t-m_{\rho}^2)$.
In particular, this subtraction eliminates $1/q^2$ and $1/q^4$
singularities corresponding to  the condensate
$\delta(t)$- and $\delta'(t)$-contributions into  $\sigma_n(t)$.
For the perturbative contribution, the subtraction procedure
removes  the $\log q^2$ terms.
If $\sigma^{PT}(t) \sim  t$
(this produces  the $q^2 \log q^2 $ contribution),
the ``correction  term'' is  given by
\begin{equation}
-\int_0^{s_{\rho}} \frac{t dt }{t+q^2} + \frac{g^{(1)}}
{q^2+m_{\rho}^2} =
-s_{\rho} +q^2 \log \left (\frac{s_{\rho}+q^2}{q^2} \right )
+\frac{g^{(1)}}{q^2+m_{\rho}^2}.
\label{eq:cort}
\end{equation}
As a result, the $- q^2 \log q^2$ term cancels the first
non-analytic term in the three-point function,
and effectively one gets $\log s_{\rho}$ instead of
$\log q^2 $ in the small-$q^2$ region.
In the  large-$q^2$ region,  where the original OPE must be valid,
the correction  terms should disappear.
Requiring that they vanish there faster than  the contribution
which they are correcting ($i.e.,$ faster than $1/q^2$)
we arrive at the relation $g^{(1)} = s_{\rho}^2/2$.
In a similar way,  if
$\sigma^{PT}(t) \sim  t^2$
(this  gives the
$q^4 \log q^2 $ term) one gets the relation
$g^{(2)} = - s_{\rho}^3/3$.

Imposing a universal $n$-independent prescription is equivalent
to the assumption  that the $y$-dependence of the
$\rho$-meson distribution amplitudes coincides
with that of the  perturbative correlators
and, furthermore, that the $\rho$-meson
contribution is dual to the  quark one,
with the standard duality interval
$s_{\rho} \approx 1.5 \, GeV^2$ obtained in ref.\cite{svz}.
In principle, one can use more elaborate  models for the
distribution amplitudes $\varphi_{\rho}^{(i)}(y)$.
As far as the requirement
of smallness of the additional terms
in the large-$q^2$ region
is fulfilled,  our results do not show
 strong sensitivity
to  particular forms of $\varphi_{\rho}^{(i)}(y)$.
However, using the local duality approximation for
$\varphi_{\rho}^{(i)}(y)$ we are able
to present  our results in a more compact and explicit
form.

After the modifications described above,
the contribution of the triangle diagram
converts into
\begin{eqnarray}
\Phi_0(M^2, q^2,Q^2) =
\frac1{\pi M^2} \int_0^{\infty} ds \, e^{-s/M^2} \frac{Q^2}{(s+Q^2)^2}
\biggl \{
 1+
 e^{y q^2/M^2}                      \nonumber \\
\left[
     2\frac{y}{M^2}
 \left(
         q^2 \ln {\frac{y(s_{\rho}+q^2)}{M^2} }
 -s_{\rho} + \frac{s_{\rho}^2}{2(m_{\rho}^2+q^2)}
         \right ) \right.
  \nonumber \\    \left.
   + \frac{y^2}{M^4}
          \left(
          q^4\ln{\frac{y(s_{\rho}+q^2)}{M^2}} - q^2 s_{\rho}
         +\frac{s_{\rho}^2}{2} -\frac{s_{\rho}^3}{3(m_{\rho}^2+q^2)}
          \right)
     \right ]
 + \ldots
\biggr \}
\label{eq:ident}
\end{eqnarray}
(here $y=s/(s+Q^2))$.
As promised, in this expression,
we can take the limit $q^2 \to 0$
  without
encountering any  non-analytic terms.
Note, that  the modified versions of
$q^2 \log q^2$ and $q^4 \log q^2$ terms
do not vanish in the $q^2 \to 0$ limit.
Finally, using the formula
\begin{eqnarray}
\int_0^{\infty}  e^{-s/M^2} g(s)  \frac{ds}{M^6} =
\int_0^{\infty}  e^{-s/M^2} ( g(0)\,\delta(s) + g'(s) ) \frac{ds}{M^4}
\nonumber \\
= \int_0^{\infty}  e^{-s/M^2} ( g'(0)\,
\delta(s) + g(0)\,\delta'(s) + g''(s) ) \frac{ds}{M^2}
\label{eq:m4tom2}
\end{eqnarray}
we can rewrite the $q^2 \to 0$ limit  of eq.(\ref{eq:ident})
in the canonical form
(\ref{eq:fph}) and determine the  relevant  spectral density
$\rho_0(s,Q^2)$.

Most of  the singular   condensate contributions of the
original sum rule (\ref{eq:SR1})
can  be interpreted in terms of the bilocals
corresponding to the simplest, one-propagator
 coefficient function.
As a result, they are subtracted by
the procedure described above.
To  factorize  the gluon condensate contribution,
we used a technique similar to that
developed for the perturbative term.
For the diagrams with  the quark condensate,
the factorization in most cases is trivial.
However, there are also terms with
the  coefficient function  formed by
three propagators (see Fig.2$b$).
In this case, the long-distance contribution is described
by the photon distribution amplitude $\phi^T(y, q^2)$
related to the
${\cal O} \sim \bar q \gamma_5 [\gamma_{\alpha}, \gamma_{\beta}]
D \ldots D q$
operators.  The  OPE
for such a ``non-diagonal'' correlator
(cf. \cite{eric}) starts with the term proportional
to the quark condensate:
$\phi^T(y, q^2) \sim  \frac{1}{q^2} \langle \bar qq
\rangle [ \delta(y) + \delta(1-y)]$,
strongly peaked at the end-points.
However, incorporating  the nonlocal condensates
to model the higher-dimension contributions
and employing a  novel technique \cite{minn}
applied earlier to a similar   non-diagonal correlator,
we found that the
distribution amplitude of the lowest-state ($i.e.,$ $\rho$-meson)
is rather narrow.
Neglecting the higher-state contributions
(whose distribution amplitudes have
oscillatory form) we obtain in the $q^2=0$ limit
\begin{eqnarray}
\Phi^{T}(M^2,Q^2) =  \frac{128\pi^2
\alpha_s \langle \bar{q}q\rangle ^2}{27   m_{\rho}^2 Q^4  M^6}
\int_0^{\infty} e^{-s/M^2} s ds \int_{s/(s+Q^2)}^1
 \frac{\varphi_{\rho}^T(y)}{y^2} dy ,
\label{eq:tensor}
\end{eqnarray}
where $\varphi_{\rho}^T(y)$ is the normalized distribution amplitude
(its zeroth $y$-moment  equals  1),
which we  model by $\varphi_{\rho}^T(y) = 6y(1-y)$.
Transforming $\Phi^{T}(M^2,Q^2)$ to the
canonical form by using eq.(\ref{eq:m4tom2})
gives the spectral density
$ \rho^{T}(s, Q^2)$.  Note that, for our
model,  $ \rho^{T}(s, Q^2)$  contains the  $(1/s)_+$ distribution.

The OPE for the three-propagator coefficient function
also produces  operators
 like $\bar q  \ldots \gamma_{\mu } D^{\mu } q$.
Naively, one would expect that such operators vanish
due to equations of motion.
However, when inserted in
a correlator, they   produce the so-called contact terms
\cite{contact}.
In our case, the contact terms give
\begin{equation}
\Phi^{C}(Q^2,M^2) = - \frac{256\pi^2
\alpha_s \langle \bar{q}q\rangle ^2}{27   Q^6  M^6}
\int_0^{\infty} e^{-s/M^2}
\left [ \ln{\frac{s+Q^2}{s}} - 2 \frac{Q^2}{s+Q^2} \right ]s ds  .
\label{eq:contact}
\end{equation}
Again, the  spectral density $ \rho^{C}(s, Q^2)$,
obtained after applying eq.(\ref{eq:m4tom2}),
 contains the  $(1/s)_+$ distribution.

Finally, there are also  configurations with the coefficient functions
given by two propagators (see Fig.2$c$)  which
correspond to  three-body $\bar q G q$-type distribution amplitudes.
Their  contribution was found to be small and,
to simplify the presentation,
 we will not include them  here.

\section {Sum rule for $F_{\gamma^* \gamma^* \pi^0 }(q^2,Q^2)$
in the  $q^2=0$ limit.}

Since  all the
 contributions, which were
singular or non-analytic in the small-$q^2$ limit
of the original sum rule, are now properly modified,
 we can take the limit $q^2=0$ and  write down our QCD sum rule
for the $\gamma \gamma^* \to \pi^0$ form factor:
\begin{eqnarray}
& \,& \pi f_{\pi} F_{\gamma \gamma^* \pi^0}(Q^2) =
\int_0^{s_0}
\left \{ %1
1 - 2 \frac{Q^2-2s}{(s+Q^2)^2}
\left (s_{\rho} - \frac{s_{\rho}^2}{2 m_{\rho}^2} \right )
\right.  \nonumber \\
&+& \left. 2\frac{Q^2-6s+3s^2/Q^2}{(s+Q^2)^4} \left (\frac{s_{\rho}^2}{2}
 - \frac{s_{\rho}^3}{3  m_{\rho}^2} \right )
\right \} %1
 e^{-s/M^2}
\frac{Q^2 ds }{(s+Q^2)^2}
 \nonumber \\
&+&\frac{\pi^2}{9}
{\langle \frac{\alpha_s}{\pi}GG \rangle}
\left \{ %2
\frac{1}{2 Q^2 M^2} + \frac{1}{Q^4}
- 2 \int_0^{s_0} e^{-s/M^2} \frac{ds }{(s+Q^2)^3}
\right \} %2
 \nonumber \\
&+&\frac{64}{27}\pi^3\alpha_s{\langle \bar{q}q\rangle}^2
\lim_{\lambda^2 \to 0}
\left \{ %3
\frac1{2Q^2 M^4}
+ \frac{12}{Q^4 m_{\rho}^2 }
\left [ %4
\log \frac{Q^2}{\lambda ^2} -2
\right.  \right. \nonumber \\
&+& \left.  \left. \int_0^{s_0} e^{-s/M^2}
\left ( %5
\frac{s^2+3sQ^2+4Q^4} {(s+Q^2)^3} - \frac1{s+\lambda ^2}
\right) ds %5
\right] %4
\right. %3
\nonumber \\
&-&
\left.  %3
\frac4{Q^6}
\left [ %6
\log \frac{Q^2}{\lambda^2} -3+
\int_0^{s_0} e^{-s/M^2}
\left (  %7
\frac{s^2+3sQ^2+6Q^4} {(s+Q^2)^3} - \frac1{s+\lambda ^2}
\right) ds %7
\right] %6
\right \} .%3
\label{eq:finsr}
\end{eqnarray}
The sum rule must be taken
in the limit $\lambda^2 \to 0$ of the parameter $\lambda ^2$
 specifying the
regularization which we used to calculate the integrals
with the $(1/s)_+$ distribution.
Furthermore, this sum rule implies that the
continuum
is modeled by an effective spectral density
$ \rho^{eff}(s, Q^2)$ rather than
by $ \rho^{PT}(s, Q^2)$, with $ \rho^{eff}(s, Q^2)$
including  all the spectral densities
which are nonzero for $s>0$, $i.e.,$ $ \rho_{0}(s, Q^2)$,
$ \rho^{T}(s, Q^2)$, $ \rho^{C}(s, Q^2)$ and
also an analogous contribution from the gluon condensate term.

Using the standard values for the condensates
 and $\rho$-meson duality interval
 $s_{\rho} = 1.5 \, GeV^2$,  \cite{svz},
we studied
the stability of the sum rule
with respect to variations of the SVZ-Borel parameter $M^2$
in the region $M^2 > 0.6 \, GeV^2$.
Good stability was observed not only for the ``canonical'' value
$s_0^{\pi} = 4 \pi^2 f_{\pi}^2 \approx  0.7 \, GeV^2$
but  also for smaller values
of $s_0$, even as small as  $0.4 \, GeV^2$.
Since our results are sensitive to  the  $s_0$-value,
we incorporated  a more detailed model for the spectral density,
treating the $A_1$-meson  as a separate resonance at
$s =1.7 \, GeV^2$,
with the continuum starting at some larger  value $s_A$.
The results obtained in this way
have good $M^2$-stability and, for $M^2 < 1.2 \, GeV^2$,
show no significant
dependence on $s_A$. Numerically, they practically
coincide with the results
obtained from the sum rule (\ref{eq:finsr})
for $s_0 = 0.7 \, GeV^2$.

\begin{figure}[t]
\mbox{
   \epsfxsize=12cm
 \epsfysize=16cm
 \epsffile{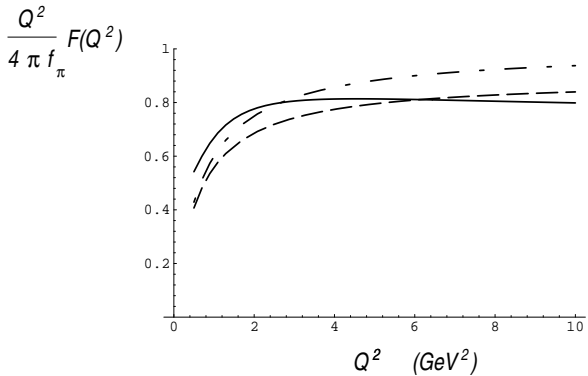}  }
  \vspace{-10cm}
{\caption{\label{fig:3}
Combination  $Q^2 F_{\gamma \gamma^* \pi^0}(Q^2)/4\pi f_{\pi}$
as calculated from the QCD sum rule
(solid line), $\rho$-pole model (dashed line)
and Brodsky-Lepage interpolation (dash-dotted line).
 }}
\end{figure}

In Fig.\ref{fig:3}, we present a curve for
$Q^2F_{\gamma \gamma^* \pi^0}(Q^2)/4\pi f_{\pi}$
calculated from eq.(\ref{eq:finsr}) for $s_0 = 0.7 \, GeV^2$
and $M^2 = 0.8\, GeV^2$.
It is rather close to the curve corresponding to the
Brodsky-Lepage interpolation
formula  $\pi f_{\pi} F_{\gamma \gamma^* \pi^0}(Q^2) =
1/(1+Q^2/4\pi^2 f_{\pi}^2)$
and to that based on the $\rho$-pole  approximation
$\pi f_{\pi} F(Q^2) = 1/(1+Q^2/m_{\rho}^2)$.
It should be noted, however,
that  the  closeness of our results to the
$\rho$-pole behaviour in the $Q^2$-channel  has nothing  to do
with the explicit use of the $\rho$-contributions
in our models for the correlators in the $q^2$-channel:
the  $Q^2$-dependence of the $\rho$-pole type  emerges
due to the fact that the pion
duality interval $s_0 \approx 0.7 \, GeV^2$
is numerically  close to $m_{\rho}^2\approx 0.6\,GeV^2$.

For $Q^2 < 3 \, GeV^2$, our curve
goes slightly  above those based on the  $\rho$-pole dominance
and BL-interpolation (which are close to the
data \cite{CELLO}). This  overshooting
is a consequence of our  assumption
that $Q^2$ can be treated as a large variable:
in some  terms, $1/Q^2$ serves  as an expansion
parameter.  Such an approximation for these terms
is invalid for small $Q^2$ and  appreciably
overestimates them for $Q^2 \sim 1 \, GeV^2$
producing  enlarged values for $F_{\gamma \gamma^* \pi^0}(Q^2)$.

In the region  $Q^2 > 3 \, GeV^2$, our curve for
$Q^2F_{\gamma \gamma^* \pi^0}(Q^2)$
 is practically  constant, supporting
 the pQCD expectation (\ref{eq:gg*pipqcd}).  The
absolute magnitude  of our prediction  gives
  $I \approx 2.4$ for the $I$-integral.
Of course, this value has some uncertainty:
it will drift if we change our models
for the photon distribution amplitudes (bilocals).
The strongest sensitivity is to the choice of
$\varphi^T_{\rho}(y)$ in the tensor contribution
(\ref{eq:tensor}). However, even rather drastic changes
in the form of  $\varphi^T_{\rho}(y)$
do not increase  our result for $I$ by
more than 20\%.  The basic reason
for this stability is that the
potentially large
$1/q^2$  factor from  the relevant contribution
in  the original sum rule (\ref{eq:SR1})
is substituted  in (\ref{eq:tensor}) by a rather small (and
non-adjustible) factor
$1/m_{\rho}^2$.

Comparing the value  $I =  2.4$ with
 $I^{as} = 3$ and $I^{CZ} = 5$, we
conclude  that our result favours
a  pion  distribution amplitude
which is narrower than the asymptotic form.
Parametrizing the width of $\varphi_{\pi}(x)$  by
 a simple model $\varphi_{\pi}(x) \sim [x(1-x)]^n$,
we get that  $I=2.4$
corresponds to $n=2.5$.
The second moment $\langle \xi^2\rangle$ ($\xi$ is
the relative fraction
$\xi = x - \bar x$)
for such a function  is 0.125.
This  low value (recall that $\langle \xi^2\rangle^{as} =0.2$
while $\langle \xi^2\rangle^{CZ} = 0.43$) agrees, however,  with
the lattice calculation \cite{lattice} and
also with the recent result \cite{minn} obtained from
the analysis of a non-diagonal correlator.

 \section{ Conclusions.}

Thus, the QCD sum rules
support the expectation that the $Q^2$-dependence
of the transition form factor $F_{\gamma \gamma^* \pi^0}(Q^2)$
is rather close to a simple interpolation
between the $Q^2 =0$ value (fixed by the ABJ anomaly)
and  the large-$Q^2$ pQCD behaviour $F(Q^2) \sim Q^{-2}$.
Moreover, the QCD sum rule approach   enables  us to calculate
the absolute normalization of the
$Q^{-2}$ term.
The value produced by the QCD sum rule
 is close to that corresponding to
the asymptotic form $\varphi_{\pi}^{as}(x) = 6 f_{\pi} x (1-x)$ of
the pion distribution amplitude.
Our curve for $F_{\gamma \gamma^* \pi^0}(Q^2)$
 is also in satisfactory  agreement with the CELLO
data \cite{CELLO} and in good agreement with
preliminary high-$Q^2$ results from CLEO \cite{CLEO}.
Hence, there is  a  very  solid evidence,
both theoretical and experimental,  that  $\varphi_{\pi}(x)$
is a rather narrow function.

\section {  Acknowledgements.}

We are
grateful to W.W. Buck, A.V. Efremov, S.V. Mikhailov and
A.P. Bakulev for useful  discussions and comments.
The work  of AR  was supported
by the US Department of Energy under contract DE-AC05-84ER40150;
the work of RR was  supported
by Russian Foundation for Fundamental
Research, Grant $N^o$ 93-02-3811 and
by International Science Foundation, Grant $N^o$ RFE300.


\begin{thebibliography}{99}
\bibitem{bl80} G.P.Lepage and S.J.Brodsky, {\it  Phys.Rev. } {\bf D22} (1980)
2157.
\bibitem{pl80}  A.V.Efremov and A.V.Radyushkin, JINR report E2-11983,
Dubna (October 1978), published in {\it Theor. Math. Phys.} {\bf 42} (1980) 97;
\\
A.V.Efremov and A.V.Radyushkin, {\it Phys.Lett. } {\bf 94B } (1980) 245.
\bibitem{blpi79} S.J.Brodsky and G.P.Lepage, {\it  Phys.Lett.} {\bf 87B} (1979)
359.
\bibitem{cz84} V.L.Chernyak and A.R.Zhitnitsky,
{\it  Phys.Reports} {\bf 112} (1984) 173.
\bibitem{cz82} V.L.Chernyak and A.R.Zhitnitsky,
{\it  Nucl. Phys.} {\bf 201} (1982) 492; {\bf 214} (1983) 547(E)
\bibitem{ABJ}    S.L.Adler, {\it Phys.Rev.} {\bf 177}, 2426 (1969);\\
            J.S.Bell, R.Jackiw,  {\it Nuovo Cim.} {\bf A60}, 47 (1967).
\bibitem{blin} S.J. Brodsky and G.P.Lepage, SLAC preprint
SLAC-PUB-2587, Stanford (1980).
\bibitem{hiroshi} H.Ito, W.W.Buck and F.Gross, {\it  Phys.Lett.}
 {\bf B287}  (1992) 23.
\bibitem{CELLO}  CELLO collaboration, H.-J.Behrend et al.,
          {\it  Z. Phys.  } {\bf C 49} (1991)401 .
\bibitem{CLEO} CLEO collaboration,  V.Savinov, hep-ex/9507005 (1995).
\bibitem{svz} M.A.Shifman, A.I.Vainshtein and V.I.Zakharov,
{\it  Nucl.Phys.} {\bf B147} (1979) 385,448.
\bibitem{NeRa83} V.A.Nesterenko and  A.V.Radyushkin,
{\it Sov.J.Nucl.Phys.} {\bf 38} (1983) 284.
\bibitem{mr} S.V.Mikhailov and A.V.Radyushkin, {\it Phys.Rev.}
 {\bf D45} (1992) 1754.
\bibitem{georgietal} R.K.Ellis, H.Georgi, M.Machacek,
H.D. Politzer and G.G. Ross,  {\it   Nucl. Phys. }
{\bf B152}  (1979) 285.
\bibitem{sterman}  G.Sterman, {\it Phys.Rev.} {\bf D17} (1978) 2773.
\bibitem{tkachev} K.G.Chetyrkin, S.G.Gorishny and F.V.Tkachov,
           {\it  Phys.Lett.} {\bf B119} (1982) 407.
\bibitem{nr84} V.A.Nesterenko and A.V.Radyushkin, {\it JETP Lett.}  {\bf 39}
(1984) 707 .
\bibitem{BelKog} V.M.Belyaev and Ya.I.Kogan, preprint ITEP-29 (1984);
            {\it Int.J. of Mod. Phys.} {\bf A8} (1993) 153
\bibitem{balyung} I.I.Balitsky and A.V.Yung,  {\it Phys.Lett.} {\bf B129}
(1983)
328.
\bibitem{xiscaling} H.Georgi and H.D.Politzer,  {\it Phys. Rev.}  {\bf D14}
(1976) 1829 .
\bibitem{eric}  A.R.Zhitnitsky,   {\it Phys.Lett.} {\bf B357} (1995) 211 .
\bibitem{minn} A.V.Radyushkin, In ``Continuous advances in QCD'',
ed. by A.V.Smilga, World Scientific (1994) p. 238;  hep-ph/9406237 .
\bibitem{contact} I.I.Balitsky, D.I.Dyakonov  and A.V.Yung,  {\it Sov. Journ.
Nucl. Phys.} {\bf 35} (1982) 761.
\bibitem{lattice} D.Daniel, R.Gupta and D.G.Richards, {\it Phys.Rev.} {\bf D43}
(1991) 3715.


\end{thebibliography}
\end{document}